\documentstyle[12pt]{article}

\def\I{{\rm I}}

\newcommand{\r}{\mbox{${\bf r}$}} 
\newcommand{\B}{{\bf B}}
\newcommand{\E}{{\bf E}}
\newcommand{\V}{{\bf v}}
\newcommand{\beq}{\begin{equation}}
\newcommand{\eeq}{\end{equation}}
\newcommand{\be}{\begin{equation}}
\newcommand{\ee}{\end{equation}}
\begin{document}

\noindent {\Large \bf Electric dipole moments, 
present and future}\footnote{Plenary talk at PANIC99, Uppsala, June 1999}

\bigskip

\noindent I.B. Khriplovich

\bigskip

\noindent Budker Institute of Nuclear Physics, 630090 Novosibirsk,
Russia

\bigskip

Upper limits on the electric dipole moments 
(EDM) of elementary particles and atoms are presented, and their 
physical implications are discussed. The bounds following 
from the neutron and atomic experiments are comparable. In particular,
they strongly constrain P odd, T even interactions.
The nuclear EDMs can be studied at ion storage rings, with 
the expected sensitivity much better than $10^{-24}\;e$ cm. It would be a 
serious progress in the studies of the CP violation.

\bigskip
\bigskip
\bigskip

\noindent {\large \bf 1. UPPER LIMITS ON ELECTRIC DIPOLE MOMENTS}

\bigskip

Up to now CP violation has been reliably observed only in
the decays of the $K^0$ mesons. Recently, indications of CP violations
were found in $B^0_d/\bar{B}^0_d \to J/\psi K^0_S$ decays.
Though the effects observed can be accomodated within the Standard Model,
their true origin still remains mysterious. 

Extremely important information on the origin of CP violation follows from
the searches for electric dipole moments of the neutron, electron and atoms.  
The EDM of a 
nondegenerate quantum-mechanical system is forbidden by time-reversal
symmetry T (and by parity conservation). T invariance and CP invariance
are equivalent, due to the CPT theorem, which is based on very strong physical
grounds. Detailed discussion of discrete symmetries (as well as of 
other problems touched upon in the talk) can be found, for
instance, in book [1].

\bigskip

\noindent {\large \bf 1.1 Elementary particles}
\label{subsec:ul}

The experimental upper limit on the neutron EDM is~[2-4]
\be\label{n}
d_{\rm n}< (6 - 10) \times 10^{-26}\ e\, {\rm cm}.
\ee
The sensitivity of these experiments can be, hopefully, improved by 
2 -- 3 orders of magnitude.

The best upper limit on the electron EDM 
\be\label{e}
d_{\rm e}<4 \times 10^{-27}\ e\, {\rm cm}
\ee
was obtained in atomic experiment with Tl [5]. Hopefully, 
this limit can be pushed well into the $10^{-28}\ e\, {\rm cm}$ 
range.

I would like to quote here one more upper limit, that on the muon
EDM [6]: 
\be\label{mu}
d_{\mu}<10^{-18}\ e\, {\rm cm}.
\ee
An experiment was recently proposed to search for the muon EDM with
the sensitivity of $10^{-24}\ e\, {\rm cm}$ [7]. We will come 
back to this proposal in Section 3.

The predictions of the Standard Model are, respectively:
\be\label{nt}
d_{\rm n} \sim 10^{-32}\;-\;10^{-31}\ e\, {\rm cm};
\ee
\be\label{et}
d_{\rm e} < 10^{-40}\ e\, {\rm cm};
\ee
\be\label{mut}
d_{\mu} < 10^{-38}\ e\, {\rm cm}.
\ee

\bigskip

\noindent {\large \bf 1.2 Atoms and nuclei}
\label{subsec:at}

The best upper limit on EDM of anything was obtained in atomic 
experiment with $^{199}$Hg [8]. The result for the dipole 
moment of this atom is
\be\label{hg}
d(^{199}{\rm Hg})<9\times 10^{-28}\ e\, {\rm cm}.
\ee
Unfortunately, due to the electrostatic screening of the nuclear EDM in
this essentially Coulomb system, the implications of the result (\ref{hg}) 
are somewhat less impressive. If one ascribes the atomic dipole 
moment to the EDM of the valence neutron in the even-odd nucleus
$^{199}$Hg, the corresponding upper limit on the neutron EDM will
be an order of magnitude worse than the direct one (\ref{n}). 

It has been demonstrated, however, that the dipole moments of {\it nuclei} 
induced by the T- and P-odd nuclear forces can be about two orders
of magnitude larger
than the dipole moment of an individual {\it nucleon} [9]. In the 
simplest approximation of the shell model, where the nuclear spin 
coincides with the total angular momentum of an odd valence nucleon,
while the other nucleons form a spherically symmetric core with the 
zero angular momentum, the effective T- and P-odd 
single-particle potential for the outer nucleon is
\begin{equation}\label{wa}
W\,=\,\frac{G}{\sqrt 2}\; \frac{\xi}{2m_p}\;\mbox{\boldmath $\sigma$}\,
\mbox{\boldmath $\nabla$}\rho(r)\,.  
\end{equation}
Here $\xi$ is a dimensionless constant
characterizing the strength of the interaction in units of the 
Fermi weak interaction constant $G$; $\mbox{\boldmath $\sigma$}$ and $\r$ 
are the spin and coordinate of the valence nucleon. Using the fact that 
the profiles of the nuclear core density $\rho(r)$ and the potential
$U(r)$ are close, one can easily find now the perturbation of the wave
function caused by the interaction (\ref{wa}). The characteristic value
of the thus induced nuclear EDM is
\be\label{N}
d_{\rm N} \sim \; 10^{-21}\,\xi \ e\, {\rm cm}.
\ee

Being interpreted in terms of the CP-odd nuclear forces, the 
experimental result (\ref{hg}) leads to the following upper limit:
\be\label{xi}
\xi< 2 \times 10^{-3}.
\ee

The Standard Model (SM) prediction for this constant is
\begin{equation}\label{smx}
\xi \sim 10^{-9}. 
\end{equation}

Thus, the theoretical predictions of the SM for dipole 
moments and CP-odd nuclear forces are about six orders of 
magnitude below the present experimental upper limits on them. 
But does this mean that the discussed experiments are of no serious
interest for elementary particle physics, that they are nothing but
mere exercises in 
precision spectroscopy? Just the opposite. It means that {\it the 
searches for electric dipole moments now, at the present level of
accuracy, are extremely sensitive to possible new physics}.  

\bigskip

\noindent {\large \bf 1.3 Beyond the Standard Model}
\label{subsec:bsm}

One could argue that the discussed experiments have ruled out more
theoretical models than any other set of experiments in the history 
of physics. Still, theoretical models of CP violation surviving up
to now, are to numerous to discuss all of them, and most of them 
have too many degrees of freedom. It is convenient therefore to proceed  
in a phenomenological way: to construct 
CP-odd quark-quark, quark-gluon and gluon-gluon operators of low dimension,
and find upper limits on the corresponding coupling constants from
the experimental results for $d_{\rm n}$ and $d(^{199}{\rm Hg})$. The analysis
performed in [10,11], has demonstrated that the limits 
on the effective CP-odd interaction operators obtained from the neutron
and atomic experiments are quite comparable. These limits  are
very impressive. All the constants are several orders of
magnitude less than the usual Fermi weak interaction constant $G$.
In particular, these limits strongly constrain some popular 
models of CP violation, such as the model of spontaneous CP violation 
in the Higgs sector, and the model of CP violation in the 
supersymmetric SO(10) model of grand unification.   

\bigskip
\bigskip
\bigskip

\noindent {\large \bf 2. T ODD, P EVEN ASIDE}

\bigskip

The EDM experiments lead also to strict upper limits on the T odd, P even 
(TOPE) interactions. 

Best direct upper limit on TOPE admixture in
nuclear forces is
\beq\label{dir}
\alpha_T < 10^{-3}.
\eeq
In any renormalizable model this admixture is at least seven orders of magnitude
smaller than (\ref{dir}). Moreover, the effect never arises to second order in 
the semiweak
coupling [12]. Therefore, searches for the TOPE interactions are
searches for new physics.

To obtain bounds on TOPE interactions from the EDM experiments, we have to 
combine this interaction with P odd electroweak correction. A simple estimate
for the thus induced neutron EDM is [13,14]
\beq 
d_{\rm n} \sim {1 \over m_p}\;(G m_{\pi}^2)\;\alpha_T < 10^{-25}\ e\, {\rm cm},
\ee
which gives
\beq\label{ld}
\alpha_T < 10^{-4}.
\eeq
Numerous elaborations on this ``long-distance'' estimate (see, for 
instance, [15]) are of a certain interest for theoretical nuclear
physics, but none of them resulted in a serious improvement over the
simple-minded result (\ref{ld}).

The true improvement is reached by going over to short-distance effects. The
corresponding contribution to the neutron EDM, due to a phenomenological 
TOPE interaction and  
the P odd electroweak one, both being combined into a two-loop diagram, results
in an extremely strong upper limit [16]:
\beq\label{sup}
\alpha_T < 10^{-12}.
\eeq

Quite recently objections were made in [17] against the approach 
of [16]. In view of the importance of the result (\ref{sup}), it 
seems appropriate to discuss these objections.

1. It is argued in [17] that our calculation ``relies on an erroneous 
result for the one-loop subgraph associated with the ABJ (chiral) anomaly''.

In fact, an identity is overlooked in [17],
\[ A^{\mu\lambda\alpha}+B^{\mu\lambda\alpha}+C^{\mu\lambda\alpha}=0\] 
(in the notations of [17]), which reduces the corresponding result 
of [17] to ours.

2. In [17], the dimensional regularization (DR) is advocated against our 
estimates with a cut-off.

However, in the discussed problem, the adopted in~[17] DR (which kills 
the powerlike divergence of the diagrams discussed)
is nothing but an accurate calculation of a small contribution of large 
distances, while in [16] the estimate is made of the dominating 
contribution of short distances.  
 
Thus, there are no reasons to doubt the validity of the result (\ref{sup}).

\bigskip
\bigskip
\bigskip

\noindent {\large \bf 3. NUCLEAR ELECTRIC DIPOLE MOMENTS\\
 AT ION STORAGE RINGS}
 
\bigskip

The various upper limits on EDMs set so far constitute a valuable 
contribution to elementary particle physics and to our knowledge of
how the Nature is arranged; the null results obtained so far are 
important. But it is only natural to think of essential progress in the 
field, of finding a positive 
result, of eventually discovering permanent electric dipole moment.
So, let me add to the above rather old stories, a new one. It should 
be started with the discussion of

\bigskip

\noindent {\large \bf 3.1 Idea of new muon EDM experiment}
\label{subsec:mul}

A new experiment was recently proposed to 
search for the muon EDM [7]. The 
intention is to use a storage ring, with muons in it 
having natural longitudinal polarization. An additional spin precession 
due to the EDM interaction with external field should be monitored by 
counting the decay electrons, their momenta being correlated with the 
muon spin, due to parity nonconservation in the muon decay.

The frequency $\mbox{\boldmath $\omega$}$ of the 
spin precession with respect to the particle momentum in external 
magnetic and electric fields, ${\B}$ and ${\E}$, is
\newpage

\beq\label{bmt}
\mbox{\boldmath $\omega$} = - {e \over m} \left[\,a {\B} - 
a\,{\gamma \over \gamma +1}\,{\V}\,({\V}{\B}) -
\left(a - {1 \over \gamma^2 - 1}\right) {\V} \times {\E}\right] 
\eeq
\[ -\,\eta\,{e \over m} \left[{\E} - 
{\gamma \over \gamma +1}\,{\V}\,({\V}{\E}) + {\V} \times {\B}\right]. \]
Here the anomalous magnetic moment $a$ is related to the $g$-factor 
as follows:
$a=g/2-1$ (for muon $a=\alpha/2\pi$); ${\V}$ is the particle velocity;
$\gamma=1/\sqrt{1 - v^2}$. The last line in this formula describes the 
precession due to the EDM $d$, the dimensionless constant $\eta$ being 
related to $d$ as follows:
\[ d = {e \over 2m}\,\eta \]
Expression (\ref{bmt}) simplifies in the obvious way for
$({\V}{\B})=({\V}{\E})=0$. Just this case is considered below.

The remarkable idea of [7] is to compensate for the usual precession in 
the vertical magnetic field ${\B}$ by the precession in a radial 
electric field ${\E}$, i.e., to choose ${\E}$ in such a way that
the first line in (\ref{bmt}) vanishes at all. Then  
the spin precession with respect to momentum is due only to the 
EDM interaction with the vertical magnetic field, and since
electric fields in a storage ring are much smaller than 
magnetic ones, it reduces to
\beq\label{edm}
\mbox{\boldmath $\omega$} =\mbox{\boldmath $\omega$}_e 
= - {e \over m}\,\eta\, {\V} \times {\B}. 
\eeq
In this way the muon spin acquires a vertical component which 
linearly grows with time. The P-odd 
correlation of the decay electron momentum with the muon spin 
leads to the 
difference between the number of electrons registered above and 
below the orbit plane.

In [7], it is stated that the limit on the muon EDM can be 
improved in the planned experiment by six orders of magnitude, 
to $10^{-24}\;e$ cm.

\bigskip

\noindent {\large \bf 3.2 Nuclear dipole moments at storage rings} 

In the same way one can search for an EDM of a polarized 
$\beta$-active nucleus in a storage ring [18]. In this case as well,
the precession of nuclear spin due to the EDM interaction can be 
monitored by the direction of the $\beta$-electron momentum. 

$\beta$-active nuclei have serious advantages as compared 
to muon. 
The life-time of a $\beta$-active nucleus can exceed by many orders 
of magnitude that of a muon. The characteristic depolarization time of 
the ion beam is also much larger than the muon life-time, which is
about $10^{-6}$ s. 
According to the estimates by I. Koop (to be published), the ion
depolarization time can reach few seconds.
Correspondingly, the angle of the rotation of nuclear spin, which is 
due to the EDM interaction and which accumulates with time, may be also
by orders of magnitude larger than that of a muon. By the same reason 
of the larger life-time, the quality of an ion beam can be made much 
better than that of a muon beam.

However, necessary conditions here are also quite serious. 

First of all, to make realistic the mentioned compensation of the 
EDM-independent spin
precession by a relatively small electric field, the effective 
nuclear $g$-factor should be 
close to 2 (as this is the case for the muon). For a 
nucleus with the total charge $Ze$, mass $Am_p$, spin $\I$, and magnetic 
moment $\mu$, the effective anomalous magnetic moment is now
\[  a = {g \over 2} - 1 = {A \over Z}\,{\mu \over 2 \I} - 1. \]
Fine-tuning of $a$ is possible in many cases by taking, 
instead of a 
bare nucleus, an ion with closed electron shells. An accurate formula
for the anomaly of an ion with the total charge $z$, is
\beq\label{a}
a = {A \over 2 z}\,{\mu \over \I} - 1.00722 + {\Delta \over A m_p}
- {z \over A}\,{m_e \over m_p}.
\eeq 
As distinct from [18], we have included here 
the correction for the atomic mass excess $\Delta$.
 
The ions which look at the moment promising from the point of view 
of the EDM
searches are presented in Tables 1, 2. The isotope data are taken
from the handbook [19]. The $\beta$-decaying excited states are 
marked in Table 1 by *. 

\begin{table}[h] 
\noindent \caption{Ion properties}
\vspace{0.4cm}
\begin{center}
\footnotesize 
\begin{tabular}{|l|c|c|c|c|c|c|c|} 
\hline
&        &             &       &          &              &   & \\ 
  & $\I^{\pi}\rightarrow \I^{\pi\;\prime}$ & $\mu$     & $z$ & 
$a\times 10^3$ & $t_{1/2}$ & $Q$ (barn)& branching \\
\hline
&        &                 &       &          &              &   &\\
$^{\;\;24}_{\;\;11}$Na  & 4$^+\rightarrow 4^+$ & 1.6903(8) & 5  & 
6.5(0.5)  & 15 h  &   & 99.944\%\\
                &   &           &    &          &         &    & \\
$^{\;\;60}_{\;\;27}$Co & 5$^+\rightarrow 4^+$  & 3.799(8) & 23 &  
- 17(2)  & 5.3 y  & 0.44  &99.925\% \\
&        &                 &       &          &              &   &\\
$^{\;\;82}_{\;\;35}$Br      & 5$^-\rightarrow 4^-$   & 1.6270(5) & 13  
& 18.0(0.3) & 35 h & 0.75& 98.5\% \\
                &   &           &    &          &         &    & \\
$^{\;\;94}_{\;\;37}$Rb & 3$^-\rightarrow 3^-$ & 1.4984(18) & 23 & 
12.5(1.2) & 2.7 s & 0.16 &30.6\% \\
                &   &           &    &          &         &    & \\
$^{110}_{\;\;47}$Ag*& 6$^+\rightarrow 5^+$  & 3.607(4) & 33 & - 6(1) 
& 250 d & 1.4  &66.8\% \\
                &   &           &    &          &         &    & \\
$^{118}_{\;\;49}$In* & 8$^-\rightarrow 7^-$ & 3.321(11) & 25 &  
- 28(3) & 8.5 s &  0.44&1.4\% \\
&        &                 &       &          &              &   & \\
$^{120}_{\;\;49}$In* & (8$^-)\rightarrow 7^-$ & 3.692(4)  & 27 & 
17(1) & 47 s & 0.53 &84.1\%\\
&        &                 &       &          &              &   & \\
$^{121}_{\;\;50}$Sn & 3/2$^+\rightarrow 5/2^+$ & 0.6978(10) & 28 & 
2.9(1.4) & 27 h & - 0.02(2)&100\%\\
  &        &            &       &          &              &   & \\
$^{125}_{\;\;51}$Sb & 7/2$^+\rightarrow 5/2^+$ & 2.630(35) & 47 & 
- 9.0(1.3) & 2.8 y & &40.3\%\\
  &        &          &       &          &              &   & \\
$^{131}_{\;\;53}$I & 7/2$^+\rightarrow 5/2^+$ & 2.742(1)  & 51  & 
- 1.9(0.4) &  8.0 d  & - 0.40&89.9\%\\
  &        &            &       &          &              &   & \\
$^{133}_{\;\;53}$I  & 7/2$^+\rightarrow 5/2^+$& 2.856(5)  & 53  & 
16(2)  & 21 h      & - 0.27&83\%\\
  &        &      &       &          &              &   & \\
$^{133}_{\;\;54}$Xe & 3/2$^+\rightarrow 5/2^+$ & 0.81340(7) & 36  
& - 6.37(9)  & 5.2 d & 0.14 &99\%\\
  &        &           &       &          &              &   & \\
$^{134}_{\;\;55}$Cs & 4$^+\rightarrow 4^+$   & 2.9937(9) & 51  &  
- 24.9(0.3)   & 2.0 y     & 0.39&70.11\%\\
 &        &            &       &          &              &   & \\
$^{136}_{\;\;55}$Cs & 5$^+\rightarrow 6^+$   & 3.711(15) & 51  &  
- 18(4)  & 13 d      & 0.22 &70.3\%\\
&        &             &       &          &              &   & \\
$^{137}_{\;\;55}$Cs & 7/2$^+\rightarrow 11/2^-$ & 2.8413(1) & 55  
& 3.0(0.1) & 30 y & 0.051&94.4\%\\
 &        &                 &       &          &          &   & \\
\hline
\end{tabular}
\end{center}
\end{table}

\newpage

\begin{table}[h]
\noindent \caption{Ion properties (continued)}
\vspace{0.4cm}
\begin{center}
\footnotesize
\begin{tabular}{|l|c|c|c|c|c|c|c|}
\hline
&        &             &       &          &              &   & \\  
 & $\I^{\pi}\rightarrow \I^{\pi\;\prime}$ & $\mu$     & $z$ & 
$a\times 10^3$ & $t_{1/2}$ & $Q$ 
(barn)& branching \\
\hline 
    &        &                 &       &          &        &   & \\                                                                      
$^{139}_{\;\;55}$Cs & 7/2$^+\rightarrow 7/2^-$ & 2.696(4) & 53 & 
2(1) & 9.3 m& - 0.075&82\%\\
&        &                 &       &          &              &   & \\
$^{141}_{\;\;55}$Cs & 7/2$^+\rightarrow 7/2^-$ & 2.438(10) & 49 & 
- 6(4) & 25 s & - 0.36&57\%\\
&        &                 &       &          &              &   & \\
$^{143}_{\;\;55}$Cs & 3/2$^+\rightarrow 5/2^-$ & 0.870(4) & 41& 
3(5) & 1.8 s & 0.47&24\%\\
&        &                 &       &          &              &   & \\
$^{140}_{\;\;57}$La & 3$^-\rightarrow 3^+$ & 0.730(15) & 17 & 
- 6$\pm$21 & 1.7 d & 0.094&44\%\\
&        &                 &       &          &              &   & \\
$^{160}_{\;\;65}$Tb& 3$^-\rightarrow 2^-$ & 1.790(7) & 47 &8(4) & 
72 d & 3.8&44.9\%\\
&        &                 &       &          &              &   &\\
$^{170}_{\;\;69}$Tm& 1$^-\rightarrow 0^+$ & 0.2476(36) & 21 & 
- 5$\pm$14 & 129 d & 0.74&99.854\%\\
&        &                 &       &          &              &   &\\
$^{177}_{\;\;71}$Lu & 7/2$^+\rightarrow 7/2^-$ & 2.239(11) & 57 & 
- 15(5) & 6.7 d & 3.4&78.6\%\\
&        &                 &       &          &              &   & \\
$^{183}_{\;\;73}$Ta & 7/2$^+\rightarrow 7/2^-$ & (+)2.36(3) & 61 & 
4$\pm$13 & 5.1 d &  &92\%\\
  &        &                 &       &          &              &   & \\
$^{196}_{\;\;79}$Au & 2$^-\rightarrow 2^+$   & 0.5906(5) & 29  &  
- 9.5(0.8) & 6.2 d & 0.81 &8\%\\
 &        &                 &       &          &              &   & \\
$^{198}_{\;\;79}$Au & 2$^-\rightarrow 2^+$   & 0.5934(4) & 29  & 
5.4(0.7) & 2.7 d & 0.68 &98.99\%\\
 &        &                 &       &          &              &   & \\
$^{203}_{\;\;80}$Hg & 5/2$^-\rightarrow 3/2^+$ & 0.84895(13)&34  
& 6.31(0.15)     & 47 d      &0.34&100\%\\
 &        &                 &       &          &           &   & \\
$^{222}_{\;\;87}$Fr & 2$^-\rightarrow 3^-$ & 0.63(1) & 35 & - 8$\pm$ 
16 & 14 m & 0.51&55\%\\
         &        &       &       &          &              &   & \\
$^{223}_{\;\;87}$Fr & 3/2$(^-)\rightarrow 3/2^-$ & 1.17(2) & 87 & 
- 7$\pm$20& 22 m & 1.2&67\%\\
 &        &                 &       &          &              &   & \\
$^{224}_{\;\;87}$Fr & 1$(^-)\rightarrow 1^-$ & 0.40(1) & 45 & 
- 11$\pm$ 25& 3.3 m & 0.52&42\%\\
&        &                 &       &          &              &   &\\
$^{242}_{\;\;95}$Am & 1$^-\rightarrow 0^+,2^+$ & 0.3879(15) &  47 
& - 8.4$\pm$ 3.9& 16 h &- 2.4&37\%,46\% \\
&        &             &       &          &              &   & \\
\hline
\end{tabular}
\end{center}
\end{table}

The errors in the values of anomalous magnetic moments $a$, presented
in Tables 1, 2, correspond to the experimental errors in values of $\mu$.
In fact, electron 
configurations, even with vanishing angular momentum $J_e$, produce a 
diamagnetic screening of nuclear magnetic moments. In most cases this 
correction, neglected here, is inessential indeed, but it is
truly large for $^{24}_{11}$Na,
changing its $a$-value from 0.0065, as presented in Table 1, to about
$\;$- 0.1.

All isotopes presented in Tables 1, 2 are $\beta^-$-active (their 
$\beta^-$ branchings are indicated in the last column). Fortunately,
many of them have allowed pure Gamow -- Teller transitions 
($|\Delta \I| = 1$) where the magnitude of the needed correlation 
between the electron momentum and the initial spin is on the order of 
unity. Few isotopes in the
tables have allowed mixed $\beta^-$-transitions ($|\Delta \I| = 0$). 
Here the magnitude of the needed asymmetry may change essentially 
from nucleus to nucleus. Obviously, for the allowed mixed transitions, 
as well as 
for forbidden transitions which are also presented in the tables, the 
values of the discussed asymmetry should be found experimentally. 

On the other hand, if a sufficiently large EDM signal can be attained, 
if the angle of the spin rotation can reach, say, a milliradian, one 
could think about an experiment with stable nuclei or nuclei of a 
large life-time. Their polarization could be measured in scattering
experiments (the idea advocated by Y. Semertzidis 
and A. Skrinsky). Among stable nuclei, the most suitable one seems to
be $^{139}_{\;\;57}$La, with $a=-0.039$ for the bare nucleus and 
$a=-0.004$ for its helium-like ion.

The nuclear polarization can be obtained, for instance, starting with
a hydrogen-like ion. For $Z \sim 50$, typical frequencies of 
hyperfine ground state transitions in these ions are close to the optical 
region. The ion can be polarized by optical pumping, and then stripped. 
A helium-like ion with polarized nucleus can be obtained from a
polarized hydrogen-like one through the electron capture.

In principle, the demand $a \ll 1$ can be softened by going over to
small velocities, $v/c \ll 1$. This would enhance the relative weight of
the compensating electric field. However, in this case one looses in the
magnitude of the EDM signal. 

But how significant would be the discussed experiments with
nuclei for elementary particle physics? 

The typical value of a nuclear EDM, as induced by CP-odd nuclear 
forces, is roughly independent of $A$ and $Z$, and can be estimated by
formula (\ref{N}). The upper limit (\ref{xi}) on $\xi$ corresponds
to the bound 
\beq
d_{\rm N} < 2 \times 10^{-24}\,e\,{\rm cm}, 
\eeq
and is at least as significant for elementary particle physics as the 
upper limit on the neutron EDM. So, even at the same sensitivity 
$10^{-24}\;e$ cm, as discussed in [7]
for muons, the experiments with nuclei would compete 
with the best present EDM results. Certainly, progress in this 
direction well deserves serious efforts. 

\section*{Acknowledgements}

I am grateful to the PANIC99 Organizing Committee for the 
support which made possible my participation in the Conference. 
Warm hospitality was extended to me at Institut f\"{u}r Theoretische
Physik, TU, Dresden, where this text was written.
The work was partly supported by
the Russian Foundation for Basic Research through Grant No. 98-02-17797
and by the Federal Program Integration--1998 through Project 
No. 274. 

\bigskip
\bigskip
\bigskip

\noindent {\large \bf REFERENCES}

\bigskip
 
\indent 1. I.B. Khriplovich and S.K. Lamoreaux,  CP Violation 
without Strangeness, Springer, 1998.\\
\indent 2. K.F. Smith {\it et al}, Phys. Lett. B, 234, (1990) 191.\\
\indent 3. I.S. Altarev {\it et al},  Phys. Lett. B, 276 (1992) 242.\\
\indent 4. P.G. Harris {\it et al},  Phys. Rev. Lett., 82 (1999) 904.\\
\indent 5. E.D. Commins, S.B. Ross, D. DeMille, and B.C. Regan, 
Phys. Rev. A, 50 (1994) 2960. \\
6. J. Baily {\it et al}, Zs. Phys. G, 4 (1978) 345.\\
7. Y.K. Semertzidis, in Proceedings of the Workshop 
on Frontier Tests of Quantum Electrodynamics and Physics of the 
Vacuum, Sandansky, Bulgaria, June 1998.\\
8. S.K. Lamoreaux {\it et al}, Phys. Rev. Lett., 59 (1987) 2275.\\
9. O.P. Sushkov, V.V. Flambaum, and I.B. Khriplovich, 
Zh. Eksp. Teor. Fiz., 87 (1984) 1521; 
[Sov. Phys. JETP, 60 (1984) 873].\\
10. V.M. Khatsymovsky, I.B.Khriplovich, and  
A.S. Yelkhovsky, Ann. Phys., 186 (1988) 1.\\
11. V.M. Khatsymovsky and I.B.Khriplovich, Phys. 
Lett. B, 296 (1992) 219.\\
12. P. Herczeg {\it et al}, to be published.\\
13. L. Wolfenstein, Nucl. Phys. B, 77 (1974) 375.\\
14. P. Herczeg, in Tests of Time Reversal Invariance in Neutron
Physics, World Scientific, Singapore, 1987.\\
15. W.C. Haxton, A. H\"{o}ring, and M. Musolf, 
Phys. Rev. D, 50 (1994) 3422. \\
16. R.S. Conti and I.B. Khriplovich,  Phys. Rev. Lett., 68 (1992) 3262.\\
17. M. Ramsey-Musolf, hep-ph/9905429.\\
18. I.B.Khriplovich, Phys. Lett. B, 444 (1998) 98.\\
 19. R.B. Firestone {\it et al}, Table of Isotopes, 
8th Edition, John Wiley, 1996.
 
\end{document}